\documentstyle[12pt,epsfig]{article}

\begin{document}

\begin{flushright}
UB/FI 98-1\\
April 1998 
\end{flushright}

\vspace{1cm}

\renewcommand{\thefootnote}{\fnsymbol{footnote}}
\begin{center}
{SOLAR MODELS AND NEUTRINO DEFICIT
\footnote{Contributed paper to the XVIII International Conference on 
Neutrino Physics and Astrophysics, Takayama, Japan, 4 - 9 June 1998}}\\
\vspace{1cm}
G. Conforto, C. Grimani, F. Martelli and F. Vetrano\\
\vspace{.5cm}
Universit\`a degli Studi, I-61029 Urbino, Italy\\
Istituto Nazionale di Fisica Nucleare, I-50125 Firenze, Italy
\end{center}
\vspace{1cm}

\begin{abstract}

The existing measurements of the solar neutrino flux are compared with 
the predictions of all models capable of reproducing the other solar observables. 
These predictions are supplemented by the hypothesis of neutrino oscillations with 
mass differences large enough to render energy-independent the depletion of the 
solar $\nu_e$ flux. 
It is concluded that the data are consistent with this hypothesis and that an 
energy-dependence of the solar neutrino deficit must be regarded as an attractive 
possibility but not as a compelling reality.

\end{abstract}

\section{Introduction \label{intro}}

In a previous paper \cite{conf} we have addressed the question of a possible 
energy-dependence of the solar-neutrino deficit coming to the conclusion that, at 
least for the time being, its existence must be regarded only as an attractive 
possibility and by no means as an established reality.

Our analysis was based on the comparison between the existing experimental data 
\cite{kir,gav,lan,fuk96} and the theoretical predictions, assumed to be well 
represented by those of the ``reference model'' of ref.~\cite{bah95}.

	The latter assumption is actually rather questionable. There exist in fact 
several solar models capable of reproducing all experimentally known facts but 
giving rise to somewhat different neutrino flux predictions \cite{cas97}. 
As different models of the same sun cannot be all simultaneously right, it follows 
that theoretical predictions can only be said to be known within indeterminations 
which are actually larger than those associated with the results of any single 
model.

In our previous analysis \cite{conf} we mentioned  this effect but did not take it 
into account. In this paper, we try to quantify this additional theoretical 
uncertainty by looking at the spread of the various predictions. The procedure 
adopted is described in section 2. 
With modified theoretical predictions and their errors, the statistical analysis 
of ref.~\cite{conf} is then repeated in section 3. Section 4 summarises our 
conclusions.

\section{Solar models \label{Somodels}}

An extensive review of solar models is presented in ref.~\cite{cas97}. We have 
restricted ourselves to ``standard'' models, i.e. to those capable of reproducing 
all solar observational results, including the helioseismology data.

	The solar neutrino event rates for the Gallium (Ga), Clorine (Cl) and 
Kamiokande (Ka) experiments predicted by the four considered models 
\cite{bah95,prof94,ric96,deg97} are reported in table 1. Within each model, the 
quoted errors reflect the uncertainties on the input parameters. 
These errors are highly correlated. However, it is only for the model of 
ref.~\cite{bah95} that the error correlation matrix is readily available 
\cite{fog95,conf}. 
Consequently, we have taken the errors of this model to be typical of and to 
apply to all this type of calculations.

\begin{table}
\caption{Solar neutrino event rates for the Clorine (Cl), Gallium (Ga) and 
Kamiokande (Ka) experiments predicted by the four ``standard'' models.}
\vspace{0.5cm}
\begin{center}
\begin{tabular}{lccc}
Experiment (Units) & Ga(SNU) & Cl(SNU) & Ka(10$^6$ cm$^{-2}$ s$^{-1}$)\\
\hline
 & & & \\
Ref.[6] & 136.8$\pm^8_7$ & 9.5$\pm^{1.2}_{1.4}$ & 6.62$\pm ^{0.93}_{1.12}$\\
Ref.[8]	& 136$\pm^7_6$	 & 8.9$\pm 1.1$	        & 6.4$\pm 0.9$\\ 
Ref.[9]	& 132.8$\pm 6.9$ & 8.5$\pm 1.1$	        & 6.3$\pm 0.9$\\
Ref.[10]& 128$\pm 6$     & 7.4$\pm 0.8$	        & 5.16$\pm 0.75$\\
\end{tabular}
\end{center}
\end{table}

Although rather similar, the central values of the rates are not identical. This 
is due to different assumptions entering in the various models. To take this 
effect into account, we have treated these results as independent determinations 
of the same quantity with a common variance. Accordingly, we have calculated the 
average values of the predicted rates and the components of their errors due to 
their spread. This procedure yields the results
\[
(133.4\pm 2.1)\, {\textrm{SNU}},\: (8.57\pm 0.45)\, {\textrm{SNU and}}\: 
(6.12\pm 0.33) \times 10^6 {\textrm{cm}}^{-2}{\textrm{s}}^{-1}
\]
for the Ga, Cl and Ka experiments respectively. The final covariance matrix on 
the average rate values is then obtained by adding in quadrature the errors 
above to the diagonal elements of the covariance matrix of the model of 
ref.~\cite{bah95}. These results are reported in table 2.

\begin{table}
\caption{Average predicted rates $P_i$ and their covariance matrix $V_{ij}$.
The convention for the indices $i,j$ is 1=Ga, 2=Cl, 3=Ka.
The two values on the diagonal elements correspond to the positive or negative 
choice of the asymmetric errors, the four values on the non-diagonal elements to 
the positive-positive, negative-positive, positive-negative and 
negative-negative choices.}
\vspace{0.5cm}
\begin{center}
\begin{tabular}{cccc}
 &$P_1$ = 133.4 SNU & $P_2$ = 8.57 SNU & $P_3$ = 
$6.12\times10^6${\textrm{cm}}$^{-2}${\textrm{s}}$^{-1}$\\
\hline
        &       &       &       \\
        &       & 6.30	& 4.81  \\
        & 68.5	& 5.51	& 5.52  \\
        & 53.5	& 7.35	& 5.79  \\
        &       & 6.43	& 5.06  \\
	&       &       &       \\
        & 6.30	&	& 1.09  \\
        & 5.51	& 1.64	& 1.27  \\						
 V =	& 7.35	& 2.16	& 1.31  \\				
        & 6.43	& 	& 1.53  \\
	&       &       &       \\			
        & 4.81	& 1.09	&       \\					
        & 5.52	& 1.27	& 0.97  \\			
        & 5.79	& 1.31	& 1.36  \\				
        & 5.06	& 1.53	&       \\
\end{tabular}
\end{center}
\end{table}				

	Fig. 1 illustrates the results of the models. The predictions from the four
solar models and their average values are shown in the Ga-Cl-Ka space. The 
ellipses centered on the stars are the $1\sigma$ contour for the model of 
ref.~\cite{bah95}, calculated by averaging the asymmetric errors. The black dot 
represents the average rate values and the smaller and larger ellipses centered on 
it are respectively the estimated uncertanties on the average rate values due to 
their differences and the $1\sigma$ contour for the average rate values (see 
table 2), calculated for averaged asymmetric errors.

\begin{figure}
\begin{center}\mbox{\epsfig{file=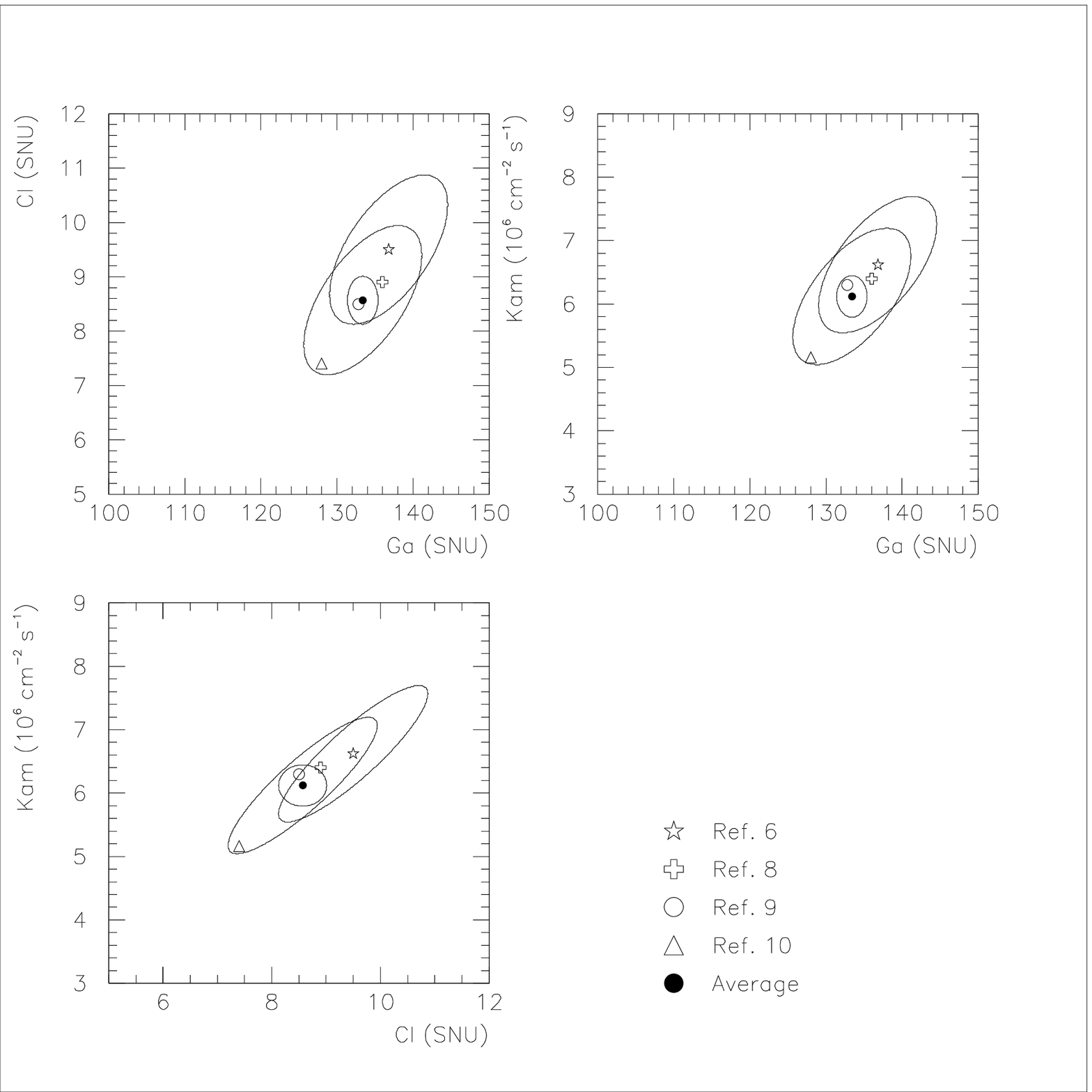,height=13cm,width=13cm}}
\end{center}
\caption{Predictions from the four solar models of 
ref.~\cite{bah95,cas97,prof94,ric96,deg97} and their average values. The 
ellipses centered on the stars are the $1\sigma$ contour for the model of 
ref.~\cite{bah95}, calculated by averaging the asymmetric errors. The black dot 
represents the average rate values and the smaller and larger ellipses centered 
on it are respectively the estimated uncertanties on the average rate values due 
to their differences and the $1\sigma$ contour for the average rate 
values (see table 2), calculated for averaged asymmetric errors.}
\end{figure}

\section{Statistical analysis \label{analysis}}
	
The experimental input data used in the analysis are shown in table 3 
\cite{nak97}.  For each result, statistical and systematic errors have been 
combined in quadrature. For the Ga and Ka experiments the weighted averages of 
the two available results have been used.

\begin{table}
\caption{Solar neutrino experimental results [12].}				
\vspace{0.3cm}
\begin{center}
\begin{tabular}{lc}
Experiment (Units) & Result\\
\hline
 & \\
Gallex (SNU)    & 69.7 $\pm 6.7 \pm ^{3.9}_{4.5}$\\
Sage (SNU)	&	73$\pm ^{10}_{11}$\\
Chlorine (SNU)	& $2.54\pm 0.14\pm 0.14$\\					
Kamiokande ($10^6{\textrm{cm}}^{-2}{\textrm{s}}^{-1}$)	&
  $2.80\pm 0.19\pm 0.33$\\	
Superkamiokande($10^6{\textrm{cm}}^{-2}{\textrm{s}}^{-1}$) &
  $2.44\pm 0.06\pm ^{0.25}_{0.09}$\\				
\end{tabular}
\end{center}
\end{table}	

Following ref.~\cite{conf}, the hypothesis of an energy-independent depletion 
of the solar $\nu_e$ flux due to oscillations is tested by studying the function

\[
\chi^{2}(F) = \Sigma_{i} \Sigma_{j} \; (e_{i} - p_{i}) \; (e_j - p_j) \;
(S^{-1})_{ij}
\]

where:
\begin{itemize}
\item  F is the common factor by which all calculated $\nu_e$ fluxes are 
reduced;

\item the indices $i$ and $j$ run over the three (Ga, Cl and Ka) experiments;

\item $e_i$ are the experimental results;

\item $p_i$ are the theoretical predictions. They are obtained from the 
$P_i$ of table 2 through the relations
\[
\begin{array}{rcl}
p_{Ga,Cl} &=& F P_{Ga,Cl} \\
p_{Ka} &=& F (1 - f) P_{Ka} + f P_{Ka}
\end{array}
\]  
where $f = 0.155$ is the fraction of the Kamiokande detection efficiency due to 
flavour-blind Neutral Currents;

\item S is the covariance matrix obtained from that of table 2 by multiplying 
the elements $V_{ij}$ with $i,j \neq 3$ by $F^2$, those with either $i$ or $j = 3$ 
by $[F^2 (1 - f) + Ff]$ and $V_{33}$ by $[F (1 - f) + f]^2$  and by adding in 
quadrature the experimental errors to the diagonal elements.
\end{itemize}

We have at first replaced all asymmetric errors by their average values. With this 
procedure, the $\chi^2$ analysis yields 
\[
F = 0.504 \pm 0.064
\]
with $\chi^2_{{min}} = 8.13$ corresponding to a Confidence Level (C.L.) 
of 1.7 $\%$.

For comparison, if the theoretical predictions and their covariance matrix are 
taken from the model of ref.~[6] alone, one obtains
\[
F = 0.437 \pm 0.056
\]
with $\chi^2_{{min}} = 14.46$ corresponding to a C.L. of 0.072 \%.

To test the sensitivity of these results to the procedure adopted in treating the 
errors, we have introduced a modification. In all operations involving two 
quantities (weighted averages of the two Ga and Ka results and evaluations of the 
$\chi^2$ terms) we have used the positive (negative) error on the first and the 
negative (positive) error on the second if the first quantity was smaller (larger) 
than the second. In this case the results are
\[
F = 0.508 \pm 0.066
\]
with $\chi^2_{{min}} = 7.61$ corresponding to a C.L. of 2.2 \%.

The results obtained by using the second procedure are shown in fig. 2. In each 
plane the two ellipses are the projections of the volumes in which the points 
representing respectively the experimental and calculated rates lie with 
68.27 \% probability.

\begin{figure}
\begin{center}\mbox{\epsfig{file=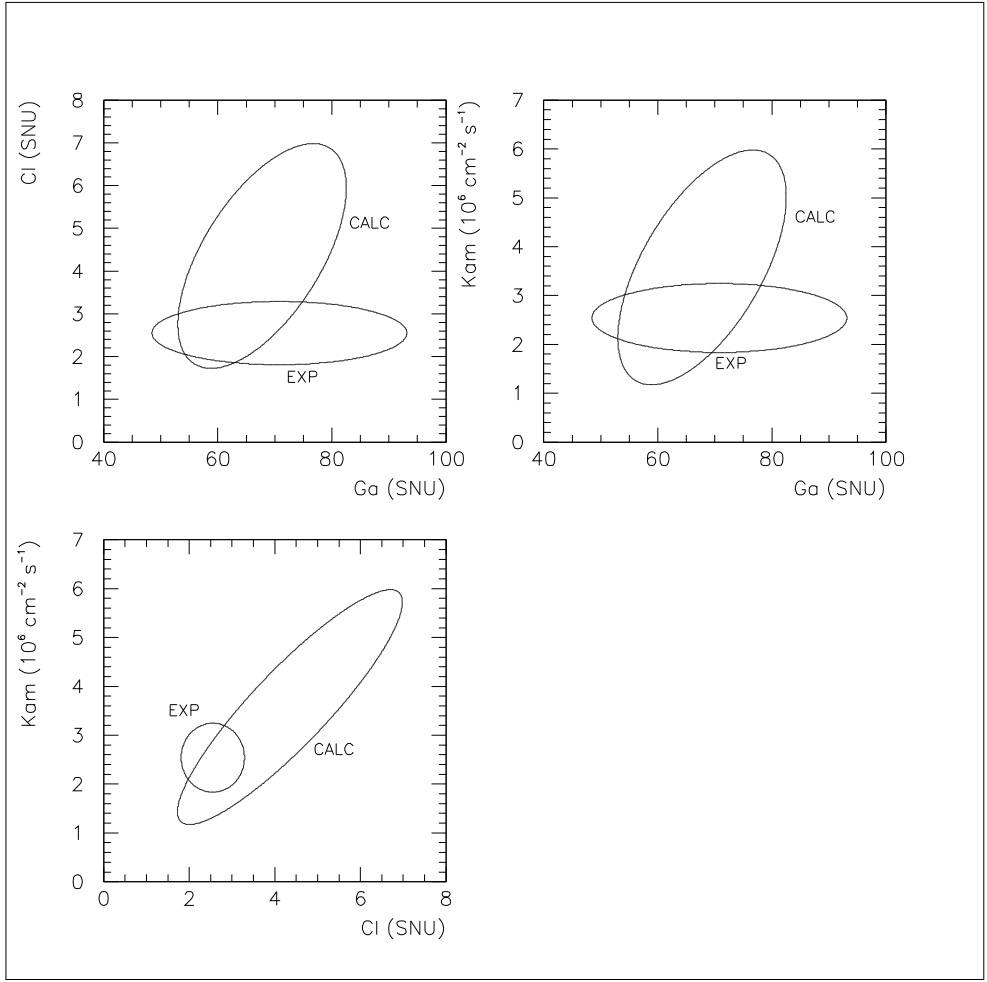,height=13cm,width=13cm}}\end{center}
\caption{Comparison between the experimental and calculated rates. The latter are 
obtained from the average rates of the models of 
ref.~\cite{bah95,cas97,prof94,ric96,deg97} supplemented by the hypothesis of 
neutrino oscillations with mass differences large enough to render 
energy-independent the depletion of the solar $\nu_e$ flux. In each plane the 
two ellipses are the projections of the volumes in which the points representing 
respectively the experimental and calculated rates lie with 68.27 \% 
probability.}
\end{figure}

\section{Conclusions \label{conc}}

Although not unacceptably low, the confidence level for an energy-independent 
depletion  of the solar $\nu_e$ flux due to oscillations obtained using the 
results of the model of ref.~[6] is admittedly rather marginal.

In the analysis presented in this paper we have used the average values of the 
predictions of four models introducing also additional few-percent uncertainties 
on them due to their spread. The confidence levels obtained are remarkably good.

In conclusion, for the time being, oscillation solutions of the solar neutrino 
problem based on an energy modulation of the solar neutrino deficit must be 
considered as perhaps suggested but certainly not compelled by the data.

\end{document}